# Securing Healthcare with Deep Learning: A CNN-Based Model for medical IoT Threat Detection

Alireza Mohammadi
*Department of Computer Engineering*
*Islamic Azad University -Kermanshah Branch*
Kermanshah, Iran
alirezamohamadi@iau.ac.ir

Hosna Ghahramani
*Department of Computer Engineering*
*Islamic Azad University- North Tehran Branch*
Tehran, Iran
haqahramani@iau.ac.ir

Seyyed Amir Asghari
*Department of Electrical and Computer Engineering, Kharazmi University,*
Tehran, Iran
asghari@khu.ac.ir

Mehdi Aminian
*Department of Computer Engineering*
*Islamic Azad University- North Tehran Branch*
Tehran, Iran
m.aminian@iau-tnb.ac.ir

*Abstract*— **The increasing integration of the Internet of Medical Things (IoMT) into healthcare systems has significantly enhanced patient care but has also introduced critical cybersecurity challenges. This paper presents a novel approach based on Convolutional Neural Networks (CNNs) for detecting cyberattacks within IoMT environments. Unlike previous studies that predominantly utilized traditional machine learning (ML) models or simpler Deep Neural Networks (DNNs), the proposed model leverages the capabilities of CNNs to effectively analyze the temporal characteristics of network traffic data. Trained and evaluated on the CICIoMT2024 dataset, which comprises 18 distinct types of cyberattacks across a range of IoMT devices, the proposed CNN model demonstrates superior performance compared to previous state-of-the-art methods, achieving a perfect accuracy of 99% in binary, categorical, and multiclass classification tasks. This performance surpasses that of conventional ML models such as Logistic Regression, AdaBoost, DNNs, and Random Forests. These findings highlight the potential of CNNs to substantially improve IoMT cybersecurity, thereby ensuring the protection and integrity of connected healthcare systems.**

## I. INTRODUCTION

The Internet of Medical Things (IoMT) represents a transformative shift in healthcare, enabling enhanced patient monitoring, diagnostics, and treatment through interconnected medical devices. However, as the adoption of IoMT technologies accelerates, so do the security concerns associated with these systems. The critical importance of IoMT security cannot be overstated; vulnerabilities in these networks can lead to dire consequences, including compromised patient safety and breaches of sensitive health information [1-2].

Traditional security methods, such as rule-based and signature-based approaches, face significant challenges in the dynamic landscape of the IoMT. These methods rely heavily on static rules and known signatures, which limit their ability to detect new and evolving threats. As a result, IoMT devices become vulnerable to novel attack vectors [3-4]. Another significant challenge is scalability. The proliferation of IoMT devices generates vast amounts of data, which traditional security methods can struggle to handle efficiently. Many IoMT devices also have limited processing power and memory, making it impractical to implement comprehensive security solutions [5-6].

Traditional security methods also lack the necessary context awareness to monitor and protect IoMT environments effectively. The IoMT landscape is highly dynamic, with devices frequently connecting and disconnecting from networks. This variability can create blind spots in security coverage and lead to false positives or negatives in threat detection [7-8]. Furthermore, Traditional security systems, requiring manual intervention, can slow threat response, and lack of skilled personnel can lead to potential security lapses in IoMT environments [9-10]. To address these challenges, Intrusion Detection Systems (IDS), deployed in IoMT networks, offer real-time threat detection, but face scalability issues and high false-positive rates, especially in the rapidly changing IoMT landscape [11-12]. Given the limitations of traditional approaches, machine learning (ML) and deep learning (DL) techniques have emerged as more effective alternatives for IoMT security. ML models can analyze large volumes of network traffic data, learn from it, and adapt to new threat patterns [13-14]. However, many conventional ML models, such as Logistic Regression and Random Forests, have demonstrated limited success in handling multiclass classification tasks, especially when distinguishing between subtle variations in attack types [15].

In this context, Convolutional Neural Networks (CNNs) offer a promising solution for intrusion detection in IoMT systems. CNNs excel at processing time-series data, which is crucial for analyzing network traffic in IoMT systems [16-17]. Their ability to automatically extract features from raw data makes them particularly effective in detecting complex, evolving threats that may be missed by traditional methods [18-19]. CNNs are well-suited for high-dimensional data and can be optimized for both binary and multiclass classification tasks [20-21]. By leveraging the power of deep learning, CNN-based intrusion detection systems can achieve high accuracy in identifying malicious activities while maintaining low false positive rates [22].

This paper presents a novel CNN-based approach for cyberattack detection in IoMT environments. By leveraging the CICIoMT2024 dataset, which contains a wide range of real-world and simulated IoMT cyberattacks, Proposed Model achieves superior performance in detecting both known and unknown threats.

## II. RELATED WORKS

The IoMT represents a highly interconnected network of medical devices, creating new avenues for patient care but also exposing critical vulnerabilities. Several studies have examined the security risks within these IoMT ecosystems. Barnett et al. (2024) highlighted the complexity of IoMT environments, where a system-of-systems design exposes networks to numerous attack vectors. Wani et al. (2023) explored specific vulnerabilities in remote patient monitoring, telemedicine, and smart medication delivery systems, pointing to real-world risks such as session hijacking and denial-of-service attacks. Kondeti and Bahsi (2024) conducted a comprehensive taxonomy of attacks on IoMT devices, including spoofing, eavesdropping, and reconnaissance [23-25]. The substantial data generated by IoMT devices, combined with the resource constraints of many medical devices, presents challenges for traditional security methods. These devices often lack the necessary computational power and memory to run comprehensive security solutions, making them easy targets for novel and sophisticated attacks [26].

The CICIoMT2024 dataset, a benchmark for cybersecurity models, contains data from 40 IoMT devices with 18 different cyberattacks. While comprehensive, the dataset poses a challenge for traditional ML models. Dadkhah et al.'s study found that while ML models perform well in binary classification, their accuracy drops significantly when trying to distinguish between all 19 attack types, often below 73%. This indicates the limitations of traditional ML in capturing the nuances of different attacks. The dataset also highlights the difficulty in distinguishing similar DDoS attacks, emphasizing the need for more advanced feature extraction and model architectures. Our proposed approach leverages CNNs to address the limitations of traditional IDS and ML models.

## III. OVERVIEW OF THE CICIOMT2024 DATASET

The CICIoMT2024 dataset is a key resource for evaluating cybersecurity solutions in the IoMT. It addresses the increasing need for secure healthcare systems, covering patient monitoring devices and remote care applications.

### A. *Key Features:*

- **Devices and Protocols:** Includes network traffic data from 40 IoMT devices (real and simulated) using common protocols like Wi-Fi, MQTT, and Bluetooth.
- **Attack Types:** Features 18 cyberattacks in five categories: DDoS, DoS, Reconnaissance, MQTT-specific, and Spoofing attacks, captured in a controlled environment for accurate security research.

### B. *Contribution:*

- **Comprehensive Coverage:** Supports security research by offering diverse devices and protocols.
- **Detailed Profiling:** Aids in detecting vulnerabilities in IoMT devices.

### C. *Class Distribution:*

- **Benign:** 230,339 instances
- **Spoofing:** 17,791 instances
- **Reconnaissance:** 926–106,603 instances
- **MQTT Attacks:** 6,877–214,952 instances
- **DoS/DDoS Attacks:**15,904–1,998,026 instances

This dataset provides an extensive foundation for analyzing cybersecurity threats in IoMT systems.

## IV. METHOD

To detect cyberattacks within IoMT environments, we have designed a CNN model optimized for analyzing time-series network traffic data. This method involves a combination of data preprocessing steps, convolutional operations, and fully connected layers to effectively classify different types of attacks. Each step of the process, from data preparation to model optimization, is crafted to ensure high performance and accuracy in identifying complex attack patterns. The overall architecture includes input layers, convolutional and pooling layers, followed by fully connected layers, as detailed below.

### A. *The Proposed CNN Model*

This section details the architecture of our proposed CNN model, designed explicitly for detecting cyberattacks within IoMT environments. We will delve into each layer's functionality, rationale behind architectural choices, and the hyperparameter selection process.

#### *1) Input Layer and Data Preprocessing:*

The input layer receives preprocessed time-series network traffic data. Each sample in the dataset is represented by a one-dimensional array of features. Before feeding the data into the model, the following preprocessing steps are applied:

- **Label Encoding and Categorical Conversion:** Attack types are encoded into numerical values and then converted to a categorical format for multi-class classification.
- **Data Standardization:** The features are standardized to have a mean of 0 and a standard deviation of 1 to improve model convergence.
- **Reshaping:** The standardized data is reshaped into a 3D tensor with dimensions (samples, features, 1) to be compatible with the 1D CNN layers.

### B. *Convolutional Layers:*

The CNN architecture consists of two 1D convolutional layers:

- First Convolutional Layer:
  - **Number of Filters:** 32
  - **Filter Size:** 3
  - **Activation Function:** ReLU (Rectified Linear Unit)
  - **Purpose:** Extracts local patterns and important features from the input data. The ReLU activation function introduces non-linearity, helping the model learn complex patterns.
- Second Convolutional Layer:
  - **Number of Filters:** 64
  - **Filter Size:** 3

- **Activation Function:** ReLU
- **Purpose:** Further refines the features extracted by the first layer, enabling the model to detect more abstract patterns.

*1) Pooling Layers:*

Following each convolutional layer, a max-pooling layer is applied:

- **Pooling Type:** MaxPooling1D
- **Pooling Size:** 2
- **Purpose:** Reduces the dimensionality of the feature maps, which helps to decrease computational cost and control overfitting by retaining only the most important features.

*2) Fully Connected Layers:*

After flattening the pooled feature maps, two dense (fully connected) layers are added:

- First Dense Layer:
  - **Number of Neurons:** 128
  - **Activation Function:** ReLU
  - **Purpose:** Combines features extracted from previous layers and enables the model to learn complex representations and decision boundaries.
- Output Layer:
  - **Number of Neurons**: Equal to the number of classes (attack types) in the dataset.
  - **Activation Function:** Softmax
  - **Purpose:** Produces a probability distribution over the classes, where each neuron's output represents the probability of the input belonging to that specific class.

### *C. Hyperparameter Selection and Optimization Strategy:*

- **Optimizer:** The Adam optimizer is chosen for its adaptive learning rate properties, which help in achieving faster convergence.
- **Loss Function:** Categorical crossentropy is used as it is well-suited for multi-class classification tasks, helping to minimize the difference between the predicted probabilities and the actual class labels.
- **Batch Size:** A batch size of 32 is selected as a trade-off between computational efficiency and the stability of the gradient estimates.
- **Epochs:** The model is trained for 10 epochs, which was determined through preliminary experimentation to ensure the model converges without overfitting.
- **Validation Strategy:** The data is split into training and validation sets to monitor the model's performance on unseen data during training, enabling early stopping if overfitting is detected.
- Evaluation Metrics: The model's performance is evaluated using accuracy, precision, recall, and F1-score, which provide a comprehensive understanding of the model's classification capabilities.

This CNN architecture (See Fig 1), with its thoughtful combination of layers and hyperparameters, is tailored to effectively detect various network attacks by learning and recognizing complex patterns in time-series data.

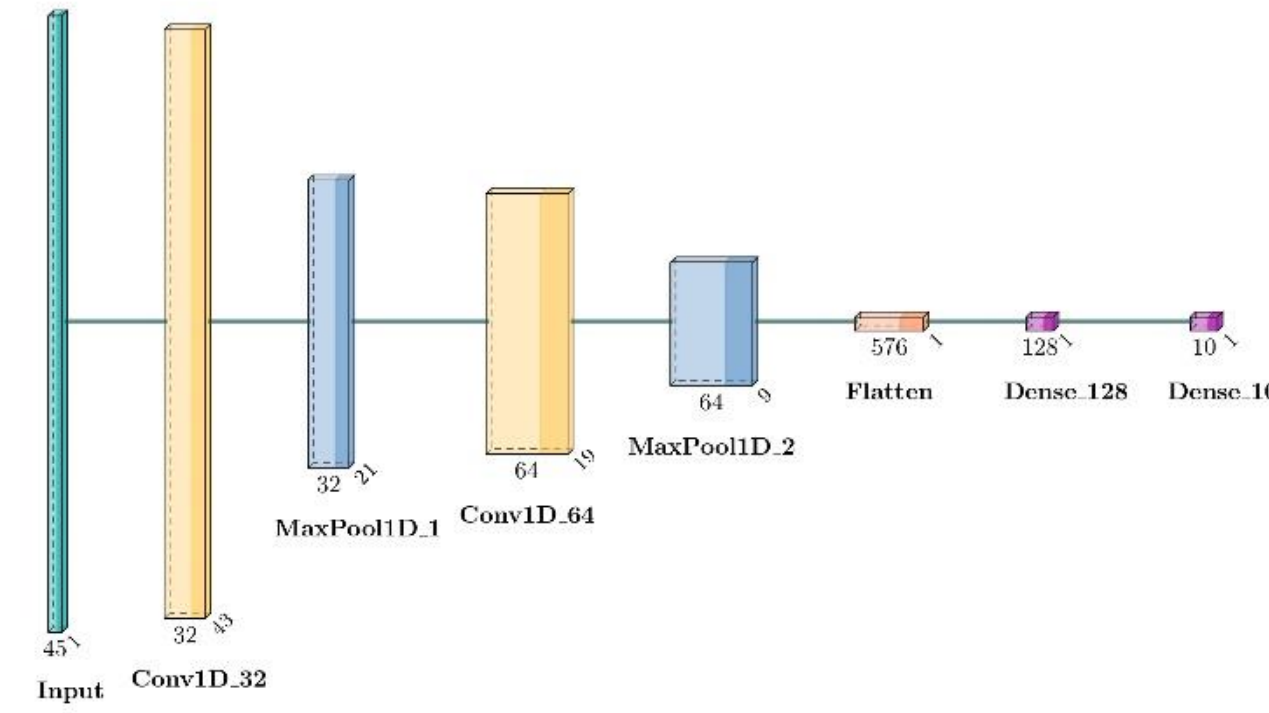

Fig 1: Architecture of the CNN model for IoMT attack classification

## V. RESULTS

This section provides a detailed comparison of our proposed model with various machine learning algorithms, including Logistic Regression, AdaBoost, Deep Neural Network, and Random Forest, across different classification tasks: binary classification, six-class classification, and 19-class classification. Table 1 presents the accuracy, precision, recall, and F1-score for each model in these three scenarios. The results clearly demonstrate the superior performance of our CNN model, especially in more complex classification tasks involving multiple attack types.

### *A. Comparative Model Performance*

Table 1 shows the performance metrics of proposed model and other ML models across different classification tasks. It highlights the effectiveness of our proposed model in handling the complexity of multi-class and categorical classifications, significantly outperforming traditional approaches like Logistic Regression and AdaBoost

Table 1: Performance Metrics of Various ML Models Across Different Classification Tasks

| Model | Classification Task | Accuracy | Precision | Recall | F1-Score |
|---|---|---|---|---|---|
| **Proposed Model** | 19 Class | 0.99 | 0.98 | 0.99 | 0.98 |
| | 2 Class | 0.99 | 0.99 | 0.99 | 0.99 |
| | 6 Class | 0.99 | 0.99 | 0.99 | 0.99 |
| **Logistic Regression [15]** | 2 Class | 0.995 | 0.959 | 0.94 | 0.946 |
| | 6 Class | 0.729 | 0.587 | 0.712 | 0.694 |
| | 19 Class | 0.727 | 0.144 | 0.471 | 0.432 |
| **AdaBoost [15]** | 2 Class | 0.996 | 0.959 | 0.961 | 0.959 |
| | 6 Class | 0.437 | 0.501 | 0.506 | 0.501 |
| | 19 Class | 0.422 | 0.141 | 0.238 | 0.141 |
| **Deep Neural Network [15]** | 2 Class | 0.996 | 0.956 | 0.948 | 0.952 |
| | 6 Class | 0.734 | 0.725 | 0.693 | 0.665 |
| | 19 Class | 0.729 | 0.649 | 0.553 | 0.522 |
| **Random Forest [15]** | 2 Class | 0.996 | 0.971 | 0.951 | 0.961 |
| | 6 Class | 0.735 | 0.735 | 0.713 | 0.676 |
| | 19 Class | 0.733 | 0.691 | 0.577 | 0.551 |

### *B. Detailed Task Evaluation*

The performance of the proposed CNN model was rigorously evaluated on the CICIoMT2024 dataset, covering binary, six-class, and 19-class classification tasks. The results for each task are detailed below, with precision, recall, and F1-score metrics presented for various cyberattack types. Figs are referred to for further clarity on the model's confusion matrices.

#### *1) Binary Classification*

In the binary classification task, distinguishing between benign and attack traffic, the model achieved near-perfect results. The accuracy was 100%, with a precision of 0.91 for benign traffic and 1.00 for attack traffic (See Fig 2).

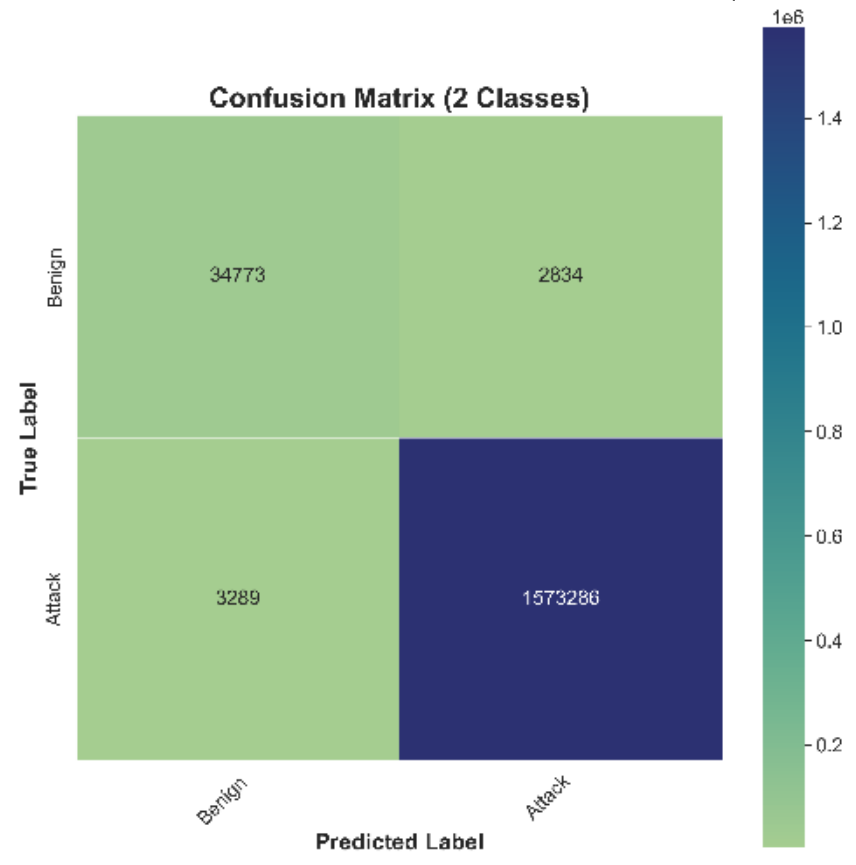

Fig 2: Confusion Matrix for Binary Classification

#### *2) Six-Class Classification*

The six-class classification task involved distinguishing between five types of Distributed Denial of Service (DDoS) and Denial of Service (DoS) attacks, alongside benign traffic. The model achieved high F1-scores of 1.00 across most attack types, with slight misclassification in categories like MQTT-DDoS-Publish Flood and MQTT-Malformed Data, as reflected in Fig 3.

The confusion matrix for this task in Fig 4 highlights the limited misclassifications, especially within certain MQTT-based attack categories.

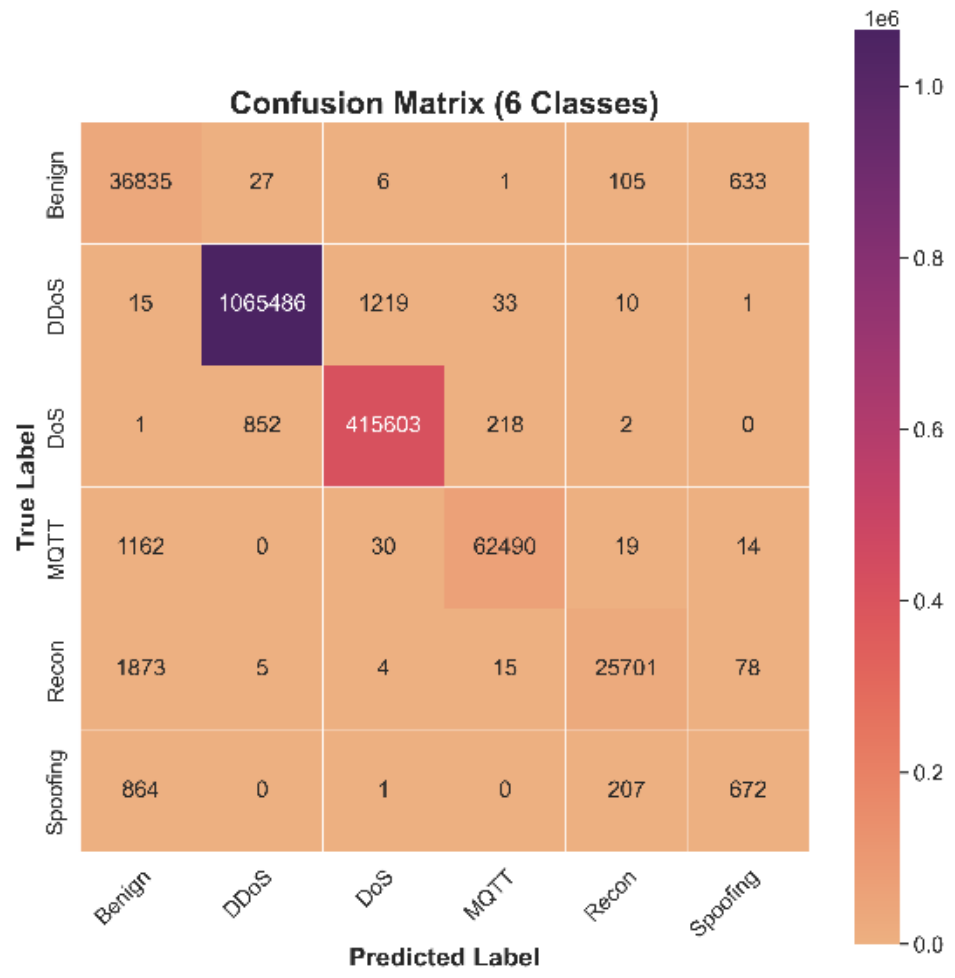

Fig 3: Confusion Matrix for Six-Class Classification

#### *3) 19-Class Classification*

For the most complex task, involving 18 different attack types, the CNN model demonstrated strong overall accuracy of 99%, as depicted in Fig 4. However, classes like Spoofing and Recon-VulScan presented more challenges for the model, with slightly lower precision and recall values compared to the DDoS and DoS classes.

The confusion matrix in Fig 4 provides detailed insight into the performance across the 19 classes.

The evaluation demonstrates that the CNN model significantly outperforms traditional ML approaches, achieving high precision and recall in both binary and multi-class classification tasks. The model effectively captures complex attack patterns across different IoMT cyberattacks. Nevertheless, there is room for improvement, particularly in the detection of low-support classes like Spoofing and MQTT-Malformed Data. Future research could focus on

**Confusion Matrix (19 Classes)**

| True Label / Predicted Label | Benign | DDoS-ICMP | DDoS-SYN | DDoS-TCP | DDoS-UDP | DoS-ICMP | DoS-SYN | DoS-TCP | DoS-UDP | MQTT-DDoS-Connect_Flood | MQTT-DDoS-Publish_Flood | MQTT-DoS-Connect_Flood | MQTT-DoS-Publish_Flood | MQTT-Malformed_Data | Recon-OS_Scan | Recon-Ping_Sweep | Recon-Port_Scan | Recon-VulScan | Spoofing |
|---|---|---|---|---|---|---|---|---|---|---|---|---|---|---|---|---|---|---|---|
| Benign | 35705 | 0 | 4 | 2 | 0 | 0 | 1 | 1 | 13 | 1 | 1 | 1 | 2 | 1 | 1 | 1 | 520 | 3 | 1350 |
| DDoS-ICMP | 2 | 349412 | 2 | 16 | 90 | 177 | 0 | 0 | 0 | 0 | 0 | 0 | 0 | 0 | 0 | 0 | 0 | 0 | 0 |
| DDoS-SYN | 4 | 1 | 171099 | 764 | 273 | 56 | 136 | 4 | 9 | 4 | 42 | 1 | 1 | 0 | 0 | 0 | 3 | 0 | 0 |
| DDoS-TCP | 1 | 12 | 1 | 182334 | 118 | 5 | 0 | 116 | 0 | 1 | 0 | 0 | 0 | 0 | 0 | 0 | 10 | 0 | 0 |
| DDoS-UDP | 2 | 7 | 1 | 72 | 361532 | 30 | 0 | 0 | 426 | 0 | 0 | 0 | 0 | 0 | 0 | 0 | 0 | 0 | 0 |
| DoS-ICMP | 1 | 181 | 0 | 12 | 3 | 98136 | 1 | 0 | 65 | 3 | 0 | 0 | 30 | 0 | 0 | 0 | 0 | 0 | 0 |
| DoS-SYN | 0 | 0 | 123 | 0 | 0 | 3 | 98467 | 1 | 0 | 0 | 1 | 0 | 0 | 0 | 0 | 0 | 0 | 0 | 0 |
| DoS-TCP | 0 | 0 | 0 | 81 | 0 | 12 | 1 | 81980 | 10 | 0 | 0 | 0 | 2 | 0 | 0 | 0 | 10 | 0 | 0 |
| DoS-UDP | 2 | 0 | 0 | 3 | 47 | 34 | 0 | 0 | 137467 | 0 | 0 | 0 | 0 | 0 | 0 | 0 | 0 | 0 | 0 |
| MQTT-DDoS-Connect_Flood | 3 | 0 | 3 | 2 | 0 | 0 | 33 | 4 | 9 | 41765 | 15 | 78 | 1 | 0 | 3 | 0 | 0 | 0 | 0 |
| MQTT-DDoS-Publish_Flood | 33 | 0 | 0 | 0 | 0 | 15 | 5 | 8 | 10 | 5 | 848 | 1 | 7490 | 0 | 0 | 0 | 1 | 0 | 0 |
| MQTT-DoS-Connect_Flood | 0 | 0 | 0 | 0 | 0 | 0 | 47 | 1 | 0 | 107 | 6 | 2968 | 1 | 0 | 1 | 0 | 0 | 0 | 0 |
| MQTT-DoS-Publish_Flood | 9 | 0 | 1 | 0 | 0 | 9 | 0 | 3 | 0 | 0 | 0 | 0 | 8483 | 0 | 0 | 0 | 0 | 0 | 0 |
| MQTT-Malformed_Data | 1376 | 0 | 0 | 0 | 1 | 0 | 0 | 0 | 1 | 9 | 2 | 0 | 0 | 322 | 24 | 0 | 12 | 0 | 0 |
| Recon-OS_Scan | 498 | 1 | 4 | 1 | 0 | 1 | 1 | 0 | 2 | 2 | 1 | 1 | 0 | 1 | 221 | 0 | 3089 | 11 | 0 |
| Recon-Ping_Sweep | 184 | 0 | 0 | 0 | 0 | 0 | 0 | 0 | 1 | 0 | 0 | 0 | 0 | 0 | 0 | 0 | 0 | 0 | 1 |
| Recon-Port_Scan | 867 | 0 | 0 | 1 | 0 | 0 | 1 | 0 | 0 | 0 | 2 | 0 | 0 | 0 | 2 | 0 | 21470 | 33 | 246 |
| Recon-VulScan | 608 | 0 | 0 | 1 | 0 | 0 | 0 | 0 | 4 | 5 | 0 | 0 | 0 | 0 | 6 | 0 | 380 | 30 | 0 |
| Spoofing | 795 | 0 | 0 | 0 | 1 | 0 | 0 | 0 | 1 | 0 | 0 | 0 | 0 | 0 | 0 | 0 | 132 | 4 | 811 |

Fig 4: Confusion Matrix for 19-Class Classification

enhancing the feature extraction techniques or refining the CNN architecture to improve detection in these more challenging categories.

## VI. DISCUSSION

The proposed CNN-based model for IoMT cyberattack detection surpasses traditional ML models like Logistic Regression, AdaBoost, DNN, and Random Forests in performance. Its advantage lies in effectively extracting complex patterns from raw network traffic data, achieving higher accuracy and F1-scores in both categorical and multiclass classification tasks. For example, in multiclass classification of 18 attack types, CNN achieves an F1-score of 0.98, while Logistic Regression scores 0.432.

A key contribution of the study is the use of the CICIoMT2024 dataset, which is tailored to healthcare-specific traffic and attacks, unlike more generic IoT datasets.

While CNN performs well, slight performance dips occur in multiclass tasks, where it struggles to distinguish closely related attacks, such as different DDoS variants. Further research into feature engineering could address this limitation.

Limitations include the model's reliance on high-quality, up-to-date training data and the computational expense of CNNs, which poses challenges for resource-limited IoMT devices. Model compression techniques and edge computing could mitigate this. The study also highlights the need for a multi-layered security approach, integrating CNN with other mechanisms like anomaly detection and access control.

For real-world applicability, the model shows promise for deployment in Network Intrusion Detection Systems (NIDS) within IoMT networks, capable of monitoring large traffic volumes and providing timely alerts. Challenges such as real-time performance, integration with existing infrastructure, and AI explainability must be addressed for practical deployment. In conclusion, the CNN model, with its focus on healthcare-specific data, brings significant novelty to IoMT security, contributing to safer, more reliable healthcare technologies.

## VII. CONCLUSION

This study introduces a CNN-based approach for cyberattack detection in IoMT environments, demonstrating significant advancements over existing machine learning benchmarks. Proposed model's success highlights the practical implications for enhancing IoMT security, particularly in its ability to maintain high accuracy and reliability in diverse and complex scenarios. Future research could focus on exploring other deep learning architectures, developing real-time detection systems, and improving the interpretability of these models, ensuring that IoMT security keeps pace with the rapidly evolving threats in the healthcare domain.

## CODE AVAILABILITY

The code used for the implementation and experimentation of this is available on GitHub at: https://github.com/alirezamohamadiam/Securing-Healthcare-with-Deep-Learning-A-CNN-Based-Model-for-medical-IoT-Threat-Detection